\documentclass[apj]{emulateapj}
\usepackage{amsmath,euscript,amsfonts,amssymb}

\def\CARMA{{\facility{CARMA}}}

\def\GBT{{\facility{GBT}}}
\def\ergphzps{\ifmmode{\mbox{erg Hz}^{-1}\mbox{ s}^{-1}}\else erg~Hz$^{-1}$~s$^{-1}$\fi}
\def\Msunpyr{\ifmmode{M_{\odot}\mbox{ yr}^{-1}}\else $M_{\odot}$ yr$^{-1}$\fi}
\def\Msun{$M_{\odot}$}

\def\mJybeam{mJy~beam$^{-1}$}
\def\kms{\ifmmode\mbox{km~s}^{-1}\else km~s$^{-1}$\fi}

\def\Kkms{K~km~s$^{-1}$}
\def\Jykms{Jy~km~s$^{-1}$}
\def\mum{$\mu$m}
\def\araa{ARA\&A}
\def\aap{A\&A}
\def\nat{Nature}

\def\mnras{MNRAS}
\def\apj{ApJ}
\def\apjs{ApJS}
\def\aj{AJ}
\def\apjl{ApJL}
\def\pasp{PASP}
\def\pasj{PASJ}
\def\hco{{\ifmmode\mathrm{HCO}^{+}\else HCO$^{+}$\fi}}
\def\hcn{{\ifmmode\mathrm{HCN}\else HCN\fi}}
\def\hc3n{{\ifmmode\mathrm{HC}_{3}\mathrm{N}\else HC$_{3}$N\fi}}

\def\chnc{{\ifmmode\mathrm{CH}_{3}\rm{NC}\else CH$_{3}$NC\fi}}
\def\h2{{\ifmmode\mathrm{H}_{2}\else H$_{2}$\fi}}

\newenvironment{figurehere*}
  {\def\@captype{figure*}}
  {}
\makeatother

\begin{document}

\title{Extended HCN and HCO$^{+}$ emission in the starburst galaxy M82}

\author{P.~Salas}
\author{G.~Galaz}
\affil{Instituto de Astrof\'isica, Facultad de F\'isica, Pontificia Universidad Cat\'olica de Chile, Av. Vicuña
Mackenna 4860, 782-0436 Macul, Santiago, Chile}
\author{D.~Salter}
\author{R.~Herrera-Camus}
\author{A.~D.~Bolatto}
\affil{Department of Astronomy and Laboratory for Millimeter-Wave Astronomy, University of Maryland, College Park, MD
20742, USA}
\author{A.~Kepley}
\affil{National Radio Astronomy Observatory, 520 Edgemont Road, Charlottesville, VA 22903-2475, USA}
\label{firstpage}

\begin{abstract}

We mapped $3$ mm continuum and line emission from the starburst galaxy M82 using the Combined Array for
Research in Millimeter-wave Astronomy. We targeted the HCN, \hco, HNC, CS and \hc3n lines, but here we focus on
the HCN and \hco\ emission. The map covers a field of $1\farcm{}2$ with a ${\approx5\arcsec}$ resolution. The HCN
and \hco\ observations are combined with single dish images. The molecular gas in M82 had been previously found to be
distributed in a molecular disk, coincident with the central starburst, and a galactic scale outflow which originates in
the central starburst. With the new short spacings-corrected maps we derive some of the properties of the dense
molecular gas in the base of the outflow. From the HCN and \hco\ ${\mbox{J}=(1-0)}$ line emission, and under the
assumptions of the gas being optically thin and in local thermodynamic equilibrium, we place lower limits to the amount
of dense molecular gas in the base of the outflow. The lower limits are ${7\times10^{6}}$~\Msun\ and
${21\times10^{6}}$~\Msun, or $\gtrsim2\%$ of the total molecular mass in the outflow. The kinematics and spatial
distribution of the dense gas outside the central starburst suggests that it is being expelled through chimneys.
Assuming a constant outflow velocity, the derived outflow rate of dense molecular gas is ${\geq0.3}$~\Msunpyr, which
would lower the starburst lifetime by $\geq5\%$. The energy required to expel this mass of dense gas is
${(1-10)\times10^{52}}$~erg.

\end{abstract}


\section{Introduction}

The low-end of the stellar mass function of galaxies is often explained as a consequence of stellar feedback
\citep[e.g.,][]{Oppenheimer2010}.
Continuous energy injection into the interstellar medium (ISM) by stellar feedback influences the way in which star
formation carries on. Numerical simulations show that continuous supernovae (SNe) explosions can end up building a
coherent galactic wind \citep*[e.g.,][]{Melioli2013}. This wind is capable of carrying ISM material from the central
regions of the galaxy to its outer regions, in some cases reaching the intergalactic medium \citep*[for a review on
galactic winds see][]{Veilleux2005}. This process can remove gas from the starburst, providing negative feedback to star
formation \citep{Springel2003,Oppenheimer2006}.
Understanding the role of galactic winds on galaxy evolution requires determining the relative contributions of
the different ISM phases to the wind \citep{Veilleux2009}. Of the different ISM phases, the relatively cold and
dense molecular gas phase is the one most directly related to star formation.
Observations of cold molecular gas in starburst-driven galactic winds are few and are only available for a couple
of objects: M82 \citep{Nakai1987,Loiseau1990,Garcia2001,Walter2002,Kepley2014}, NGC253 \citep{Bolatto2013} and NGC3628
\citep{Tsai2012}. However, these suggest that the relatively cool molecular gas phase is an important fraction of the
wind mass budget.

% \noindent
To study the relative importance of the different ISM components in an outflow we obtained maps of dense gas in one of
the closest starburst galaxies with a known starburst-driven wind, M82 \citep[$D=3.4\pm0.2$~Mpc,][]{Dalcanton2009}.
Its central starburst covers a region $\sim1$~kpc in diameter \citep*{OConnell1978}, and hosts an extreme environment
as suggested by its high star formation rate of $\sim9$~\Msunpyr \citep{Strickland2004} and infrared luminosity
$L_{\mathrm{IR}}\approx5.6\times10^{10}$~L$_{\sun}$ \citep{Sanders2003}.
Its proximity and high star formation rate have made it a target for observations covering the whole electromagnetic
spectrum.
This has permitted to draw a picture of the starburst in great detail. The starburst-driven wind in M82 was first
observed by \citet*{Lynds1963}. The outflow presents a bipolar structure, observed in X-rays
\citep*{Bregman1995,Strickland1997,Stevens2003}, optical emission lines \citep*{Shopbell1998}, CO
\citep{Walter2002}, polycyclic aromatic hydrocarbons \citep[PAHs,][]{Engelbracht2006,Yamagishi2012}, warm H$_{2}$
\citep{Veilleux2009}, dust \citep{Roussel2010}, forbidden lines in the IR \citep{Contursi2013} and dense molecular gas
\citep{Kepley2014}. These observations have shown that the different ISM phases coexist in the wind, with the cold
molecular component, traced by CO, dominating the mass and kinetic energy budget of the wind.

% \noindent
In this work we present new continuum and line emission observations of M82 obtained with the Combined Array for
Research in Millimeter-wave Astronomy (CARMA). Among them are HCN and \hco\ observations with high sensitivity and
spatial resolution. The maps of {bf \citet{Kepley2014}} were the first to reveal \hco\ and HCN emission associated with
the starburst-driven outflow in M82 thanks to the high sensitivity of the Green Bank Telescope (GBT). We combine our HCN
and \hco\ maps with the GBT data to recover the missing large scale flux resolved out by \CARMA. The use of new CARMA
data along the previously published GBT observations results in maps with a factor of $2$ better angular
resolution which enables to study the dense gas in the galactic wind with greater detail.

% \noindent
In \S\ref{sec:obs} we present the new CARMA observations and how these were combined with the GBT observations.
Then in \S\ref{sec:results} we determine the amount of dense gas present in the outflow from the resulting maps.
Finally, in \S\ref{sec:summary} we give a summary of our main findings.

\section{Observations and data reduction}
\label{sec:obs}

\subsection{CARMA observations}
We observed M82 in the $3$~mm band using \CARMA\footnote{Support for CARMA construction was derived from the states of
California, Illinois, and Maryland, the James S. McDonnell Foundation, the Gordon and Betty Moore Foundation, the
Kenneth T. and Eileen L. Norris Foundation, the University of Chicago, the Associates of the California Institute of
Technology, and the National Science Foundation. Ongoing CARMA development and operations are supported by the National
Science Foundation under a cooperative agreement, and by the CARMA partner universities.}
in its D configuration (baselines $11$-$150$ m) during April and May 2012. We observed a 7 point mosaic centered on
${\alpha=09^{\mathrm{h}}55^{\mathrm{m}}53.00^{\mathrm{s}}}$, ${\delta=+69\arcdeg40\arcmin47\arcsec}$ (J2000),
the radio center of this galaxy. This mosaic covers a circular region of $\sim1\farcm{}2$ in diameter.
The correlator was tuned to have $16\times250$~MHz windows with a $3.125$~MHz spectral resolution.
This corresponds to a velocity coverage of $\sim770$~\kms\ with $10.5$~\kms\ resolution. Eleven windows
were used to measure the continuum emission and the remaining $5$ were centered over the molecular lines shown in
Table~\ref{tab:obs}. Observations were reduced using \emph{MIRIAD} data reduction software \citep{Sault1995}. Mars was
used as a primary flux calibrator.
Passband solutions were derived by observing 0927+390, and gain solutions observing 0836+710 (or 0841+708 by its B1950
name). A linear baseline was removed from the windows with spectral lines. The baseline was determined from the
line free windows and channels. The average synthesized beam of the final cubes is $\sim5\farcs2\times4\farcs9$
($6\arcsec\approx102$~pc) and the average channel rms is $44$~mK ($\approx7.4$~\mJybeam). A summary of the cube
properties is presented in Table~\ref{tab:obs}.

\subsection{Short-spacings correction}
The shortest baseline of $11$~m makes the telescope blind to structures larger than $\sim50\arcsec$. To correct for
the negative bowl in visibility amplitude produced by the lack of short spacings we used \GBT\ HCN and \hco\
observations \citep{Kepley2014}. Images were combined in spatial frequency space using \emph{MIRIAD}'s
\emph{IMMERGE}. This method yields similar results to other missing short-spacings correction methods
\citep{Stanimirovic2002}. The flux scaling between GBT and CARMA data was determined in an annulus of $4-33$~k$\lambda$,
corresponding roughly to the dish size of the GBT and the shortest baseline in the interferometer array. Flux scaling
factors were $0.376$ and $0.4$ for HCN and \hco\ respectively.
The short spacing corrected channel rms is $44$~mK ($=7.2$~\mJybeam) for HCN and $42$~mK ($=7.0$~\mJybeam) for \hco.
The integrated line brightness of the short-spacing corrected maps is different from those presented in
\citet{Nguyen1989} by $2\%$ for HCN and $13\%$ for \hco.

\noindent
The remaining observations presented in this work, $3$~mm continuum, HNC, CS and \hc3n, were not corrected for
missing short-spacings.

\setlength{\tabcolsep}{3pt}
\begin{deluxetable}{lcccc}
\tabletypesize{\scriptsize}
\tablecaption{CARMA observations summary\label{tab:obs}}
% \tablewidth{-0.5pt}
\tablehead{
\colhead{Molecule} & \colhead{Frequency\tablenotemark{a}} & \colhead{Obs.
beam} & \colhead{\it{rms}} & \colhead{$n_{\mathrm{crit}}$\tablenotemark{b}}\\%& \colhead{Refs.\tablenotemark{b}} \\
\colhead{} & \colhead{(GHz)} & \colhead{($\arcsec\times\arcsec$)} &
\colhead{(mK)} & \colhead{($\times10^{5}$ cm$^{-3}$)}
}
\startdata
% \multicolumn{5}{c}{Detections}\\
% \hline\\ [-1.5ex]
HCN($1-0$)           & $88.63$ & $5.5\times5.2$ & $43$ & $1.9\times10^{6}$\\%& (1),(4)\\
\hco($1-0$)          & $89.18$ & $5.5\times5.2$ & $48$ & $2.1\times10^{5}$\\%& (1),(4),(5)\\
HNC($1-0$)           & $90.66$ & $5.3\times4.9$ & $43$ & $3.5\times10^{5}$\\%& (2)\\
\hc3n($10-9$)        & $90.98$ & $5.4\times5.0$ & $40$ & $3.6\times10^{6}$\tablenotemark{c}\\%&\\
CS($2-1$)            & $97.98$ & $4.9\times4.7$ & $40$ & $2.7\times10^{4}$%& (1),(6)\\
% \hline \\ [-1.5ex]
% \multicolumn{5}{c}{Non-detections}\\
% \hline \\ [-1.5ex]
% \hc3n($11-10$)       & $100.8$ & $4.9\times4.6$ & $47$ & \nodata\\%&\\
% SO($3-2$)            & $99.29$ & $4.9\times4.6$ & $50$ & \nodata\\%&\\
% \choh($2_{k}-1_{k}$) & $97.58$ & $4.9\times4.7$ & $41$ & \nodata\\%& (3)\\
% \chnc($5-4$)         & $100.5$ & $4.9\times4.6$ & $45$ & \nodata
% \ch($63/2-61/2$)     & $87.97$ & $5.5\times5.2$ & $39$ & \nodata\\%&\\
\enddata
\tablenotetext{a}{Spectral line frequencies were taken from the Spectral Line Atlas of Interstellar Molecules (SLAIM)
(Available at http://www.splatalogue.net). \citep[F. J. Lovas, private communication,][]{Remijan2007}}
\tablenotetext{b}{Critical density $n_{crit}\approx A_{u\ell}/\sum_{i}\langle v\sigma_{ui}\rangle$ where
$A_{u\ell}$ is the Einstein coefficient and $\langle v\sigma_{u\ell}\rangle$ is the collisional rate. Critical
densities were computed for $T=50$~K. Collisional rates and Einstein coefficients were taken from the Leiden
Atomic and Molecular Data base \citep[LAMDA,][]{Schoier2005}.}
\tablenotetext{c}{Collisional rates not available for \hc3n. This value was taken from \citet{Aladro2011a} which uses
$T=90$ K.}
% \tablenotetext{b}{Previous observations of the same transition in M82. (1):\citet{Nguyen1989};
% (2):\citet{Huettemeister1995}; (3):\citet{Martin2006}; (4):\citet{Henkel1998}; (5):\citet{Seaquist1998};
% (6):\citet{Bayet2008}}
\end{deluxetable}

\section{Results}
\label{sec:results}

\subsection{3 mm continuum}

The $3$~mm continuum emission from the central region of M82 is dominated by optically thin free-free emission
\citep*{Carlstrom1991}, which is produced by ionized gas ($\sim10^{4}$~K) present in H$^{+}$ regions . This
makes the $3$~mm continuum emission from M82 a good tracer of sites of recent star formation. Hence, the extent of the
$3$~mm continuum should be similar to that of the central starburst.

\begin{figure}
\includegraphics[width=1\linewidth]{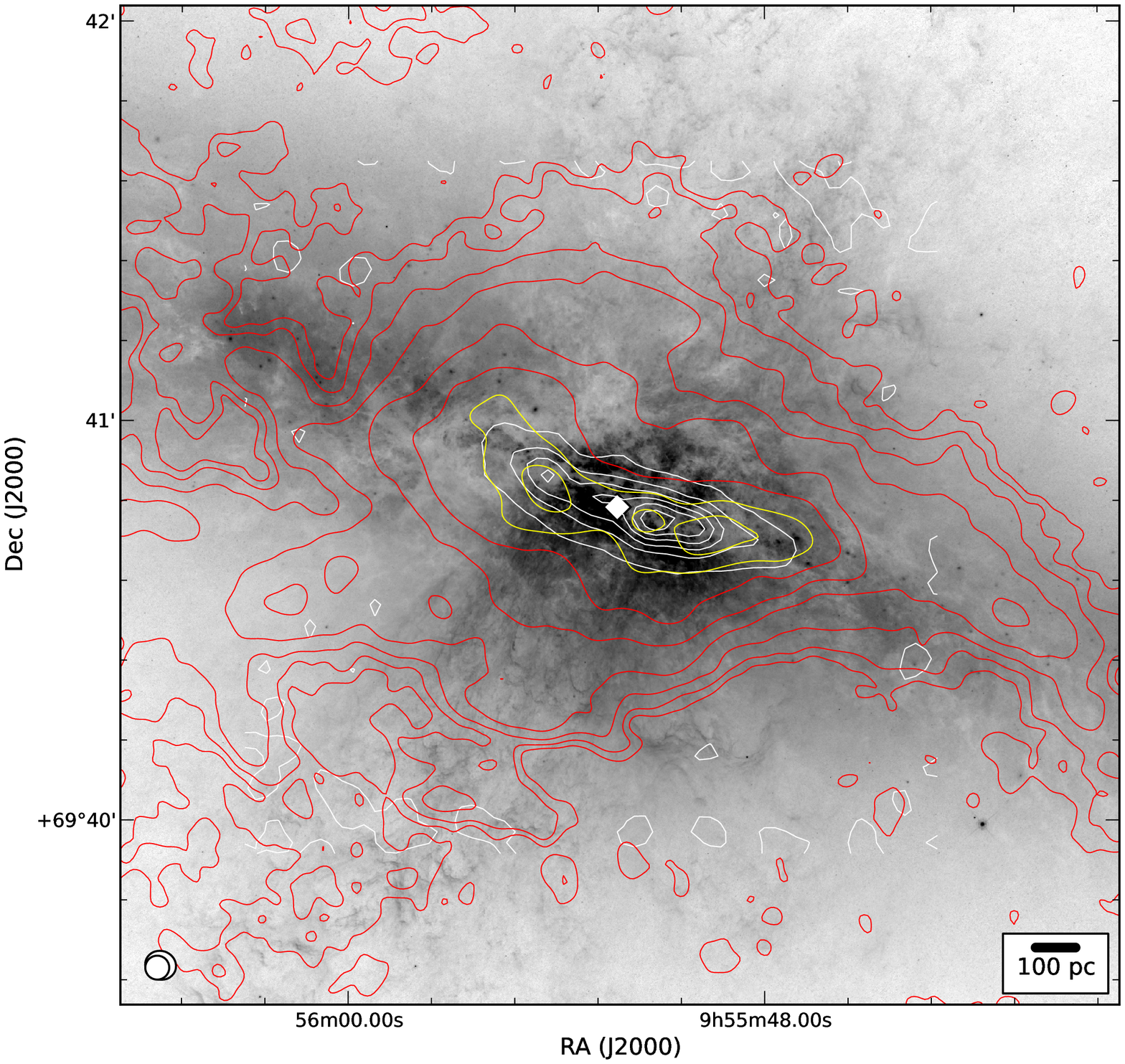}
\caption{\label{fig:cont}
3 mm continuum emission from the central region of M82 (white contours) overlaid on a H$\alpha$ emission image
from the HST/ACS \citep[inverted logarithmic gray scale,][]{Mutchler2007} and CO $\mbox{J}=1-0$ moment $0$ map
\citep[red contours: outflow; yellow contours: molecular disk][]{Walter2002}. The white diamond shows the location
of the galactic center as defined by the $2.2$~\mum\ peak \citep{Dietz1989}.
The continuum contours start at $3\sigma$, with $\sigma=21$~mK. Each increment is $6\sigma$.
The CARMA beam of the observations presented in this work is shown as the large white circle on the lower left corner.
The smaller circle shows the beam of the CO observations \citep[$3\farcs6\times3\farcs6$,][]{Walter2002}
}
\end{figure}

% \noindent
The new $3$~mm continuum map obtained with CARMA is shown in {Fig.~\ref{fig:cont}}. In the same figure we show the
H$\alpha$ emission from M82 as observed with the Hubble Space Telescope Advanced Camera for Surveys
\citep[HST/ACS,][]{Mutchler2007}, and the reference CO($1-0$) map of \citet{Walter2002}. The CO map was obtained
with a combination of Owen Valley Radio Observatory plus IRAM $30$~m data.
A visual comparison between the HST image and the $3$~mm continuum from this work (Fig.~\ref{fig:cont}) reveals
that the peaks of the $3$~mm continuum correspond to regions of high dust opacity in the H$\alpha$ image. The continuum
disk has a minor axis of $\approx7\arcsec$, similar to the number density scale height of the stellar clusters in M82
\citep*[$\approx9\arcsec$,][]{Lim2013}. The major axis of the continuum disk is of $\approx30\arcsec$. The extent
of the continuum is also similar to that of the molecular disk as observed in the CO map. This motivates us to define
the spatial extent of the molecular disk as a combination of the continuum disk presented here, and the CO
molecular disk \citep{Walter2002}.

% \noindent
The continuum map is similar to that found in previous studies
\citep{Carlstrom1989,Carlstrom1991,Brouillet1993,Neininger1998}.
In their work \citet{Brouillet1993} report a total flux of $\approx0.45$~Jy for the South-West portion of the galaxy. In
our maps the total integrated flux over the disk is ${\approx0.50\pm0.03}$~Jy. The total continuum flux is in agreement
with the single dish flux reported by \citet*[][$\approx0.54\pm0.08$~Jy]{Jura1978}.

\subsection{HCN and \hco line emission}

\begin{figure*}
\includegraphics[width=1\linewidth]{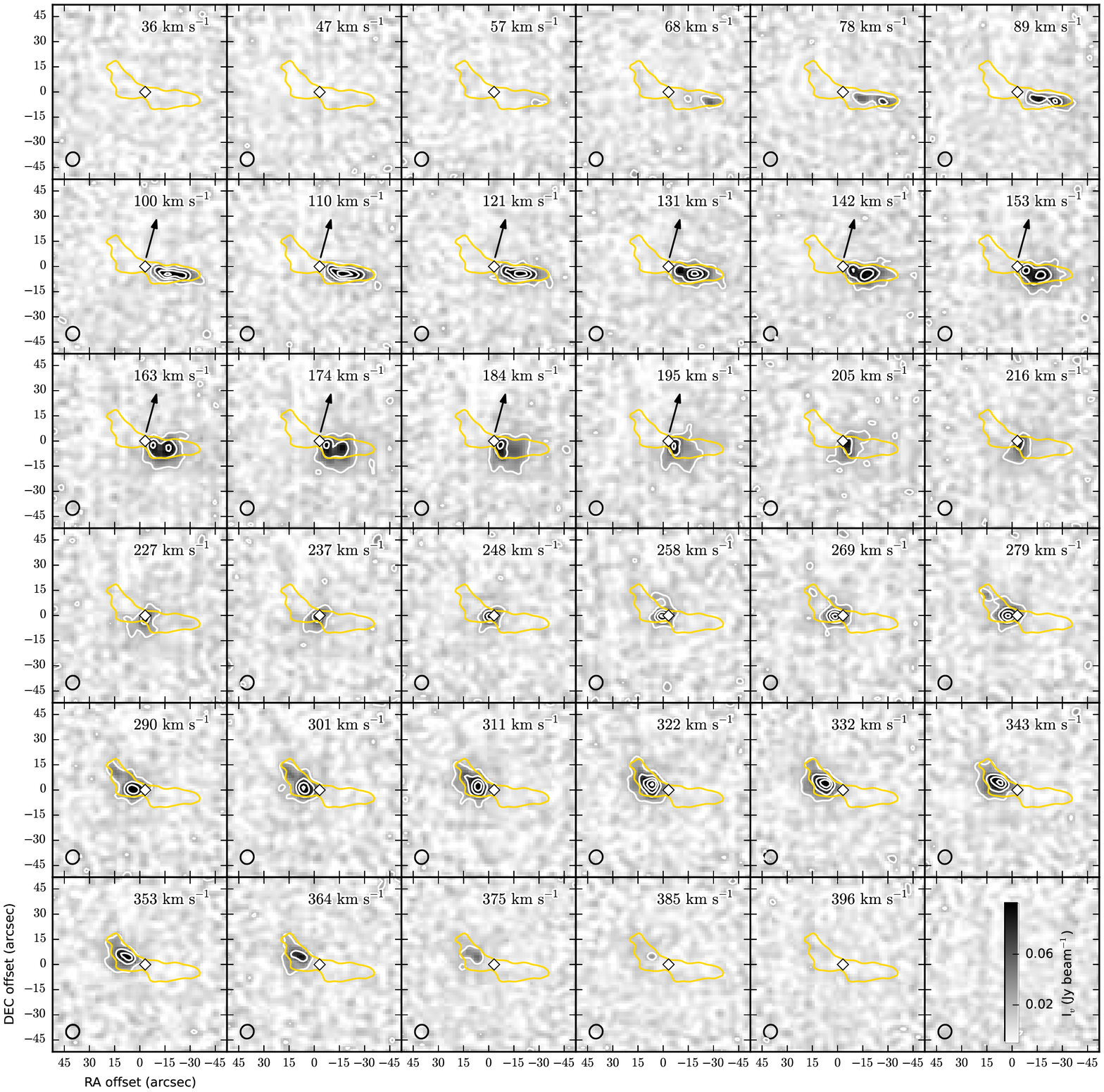}
\caption{HCN channel map. The radio velocity with respect to the local standard of rest is shown in the upper right
corner of each panel. The white diamond marks the position of the galaxy center \citep[$2.2$~\mum\ peak][]{Dietz1989},
and the arrow in some of the panels marks the location of the SiO($2-1$) chimney \citep{Garcia2001}.
The interferometer beam is shown in the lower left panel. Contours start at $3\sigma$ with $12\sigma$
increments. $1\sigma=7.1$~mJy~beam$^{-1}=43$~mK.}
\label{fig:hcn_chmap}
\end{figure*}

\begin{figure*}
\includegraphics[width=1\linewidth]{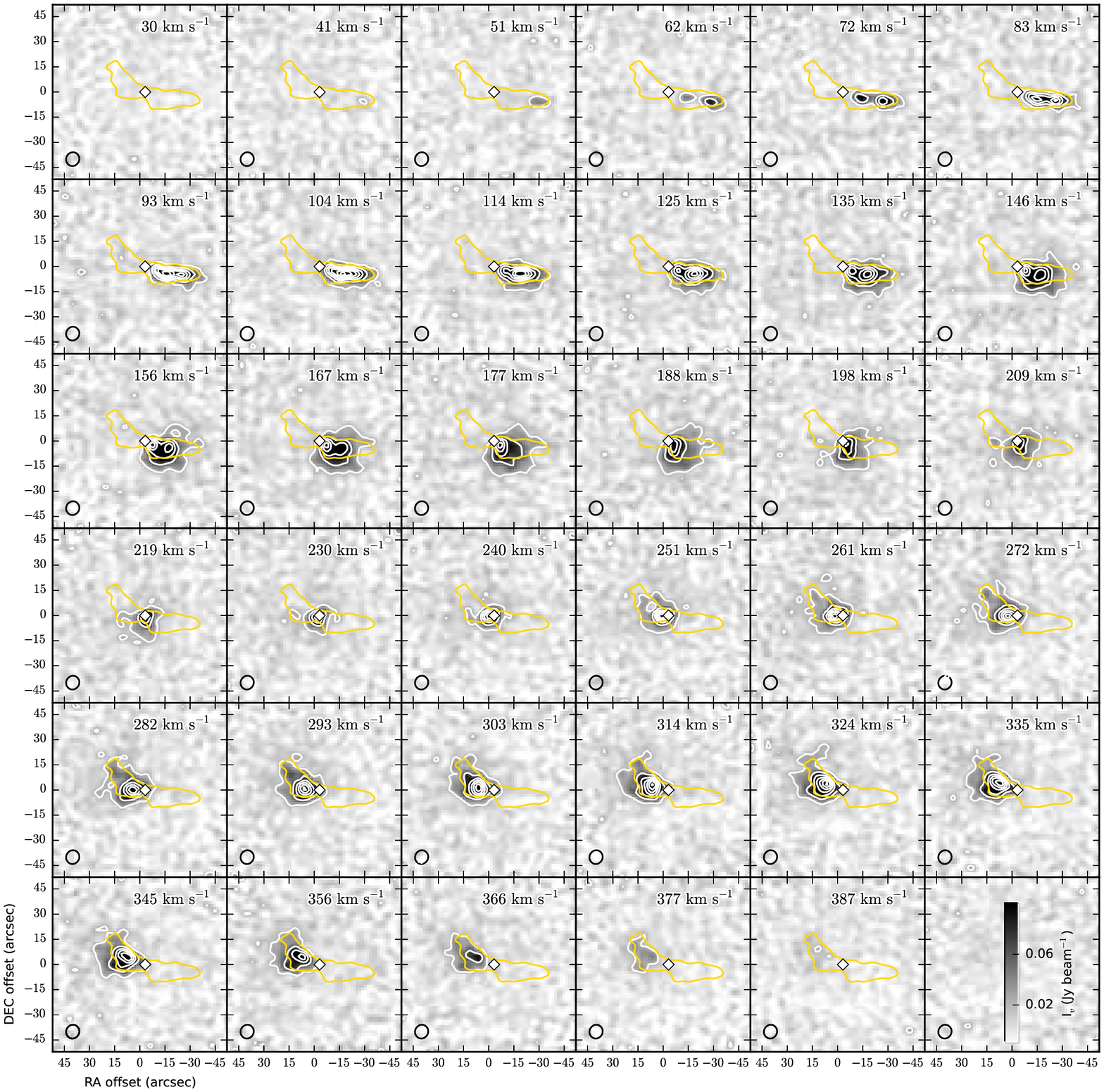}
\caption{\hco\ channel map. The radio velocity with respect to the local standard of rest is shown in the upper right
corner of each panel. The white diamond marks the position of the galaxy center \citep[$2.2$~\mum\ peak][]{Dietz1989}.
The interferometer beam is shown in the lower left panel. Contours start at $3\sigma$ with $12\sigma$
increments. $1\sigma=7.0$~mJy~beam$^{-1}=43$~mK.}
\label{fig:hco_chmap}
\end{figure*}

Channel maps corrected for missing short spacings are presented in Figs.~\ref{fig:hcn_chmap} and \ref{fig:hco_chmap}.
These show the location of extraplanar emission extending below and above the starburst disk. For HCN, in the velocity
range {$121-343$~\kms} emission extends above and below the molecular disk. The situation is similar for \hco, for
which in the velocity range {$114-366$~\kms} we see emission more extended than the molecular disk along its minor
axis. This emission comes from regions near the start of the outflow as observed in CO and H$\alpha$ (see
Fig.~\ref{fig:cont}).
This suggests that the extraplanar emission extending above and below the molecular disk might be
entrained in the galactic wind. The extraplanar \hco\ emission had been previously associated with the galactic wind in
M82 \citep{Kepley2014}. Here we extend this association to the HCN emission as well.

% \noindent
The molecular disk spans a velocity range ${52-387}$~\kms, and its kinematics have been discussed in depth by
\citet{Brouillet1993} for HCN and \citet{Seaquist1998} for \hco. The extraplanar emission follows the rotation of the
molecular disk. We find no observable counterpart to the chimney traced by SiO($2-1$) \citep{Garcia2001} in the
channel maps. The SiO chimney appears at a velocity of $150$~\kms to the East of M82 center, that is blueshifted with
respect to the disk velocity at the base of the chimney.

\begin{figure*}
\includegraphics[width=\linewidth,angle=0]{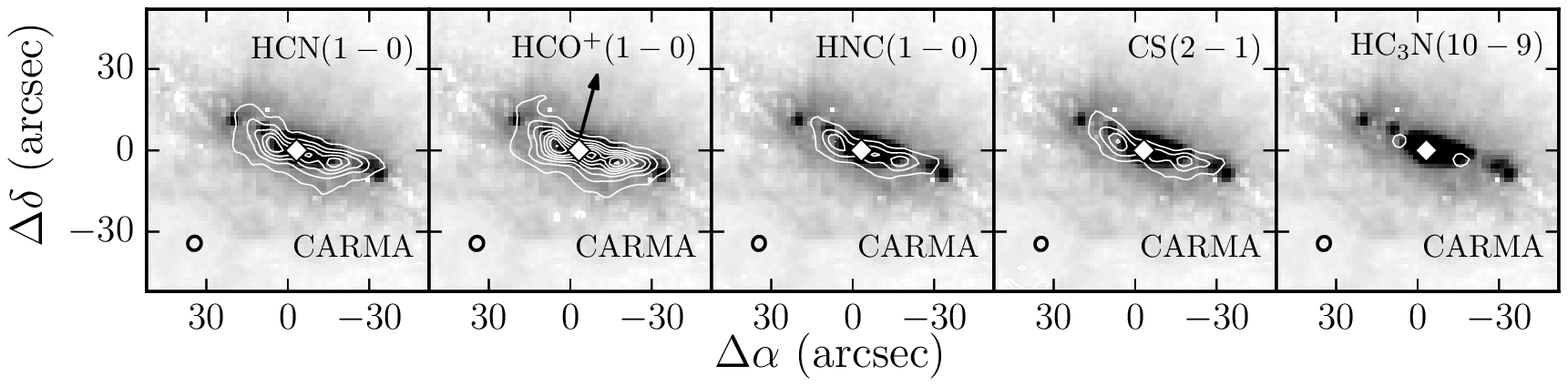}
\caption{\label{fig:mom0_carma}
Integrated intensity (moment $0$) maps of line emission (solid contours) overlaid on a map of \h2 $v=1-0$ S(1)
emission \citep[inverted gray scale,][]{Veilleux2009}. The line emission maps were obtained with the CARMA telescope
and have not been corrected for missing short spacings. The arrow on the \hco\ map shows the location of the SiO
chimney \citep{Garcia2001}.
Contours start at $3\sigma$ and increment by $6\sigma$, with $\sigma$ the moment $0$ map rms.
$\sigma_{\mathrm{HCN}}\approx2.5$~\Kkms, $\sigma_{\hco}\approx2.7$~\Kkms, $\sigma_{\mathrm{HNC}}\approx2.5$~\Kkms,
$\sigma_{\mathrm{CS}}\approx2.3$~\Kkms, $\sigma_{\mathrm{\hc3n}}\approx1.8$~\Kkms. The CARMA beam is shown on the lower
left corner of each panel.
}
\end{figure*}

\begin{figure*}
\includegraphics[width=1\linewidth,angle=0]{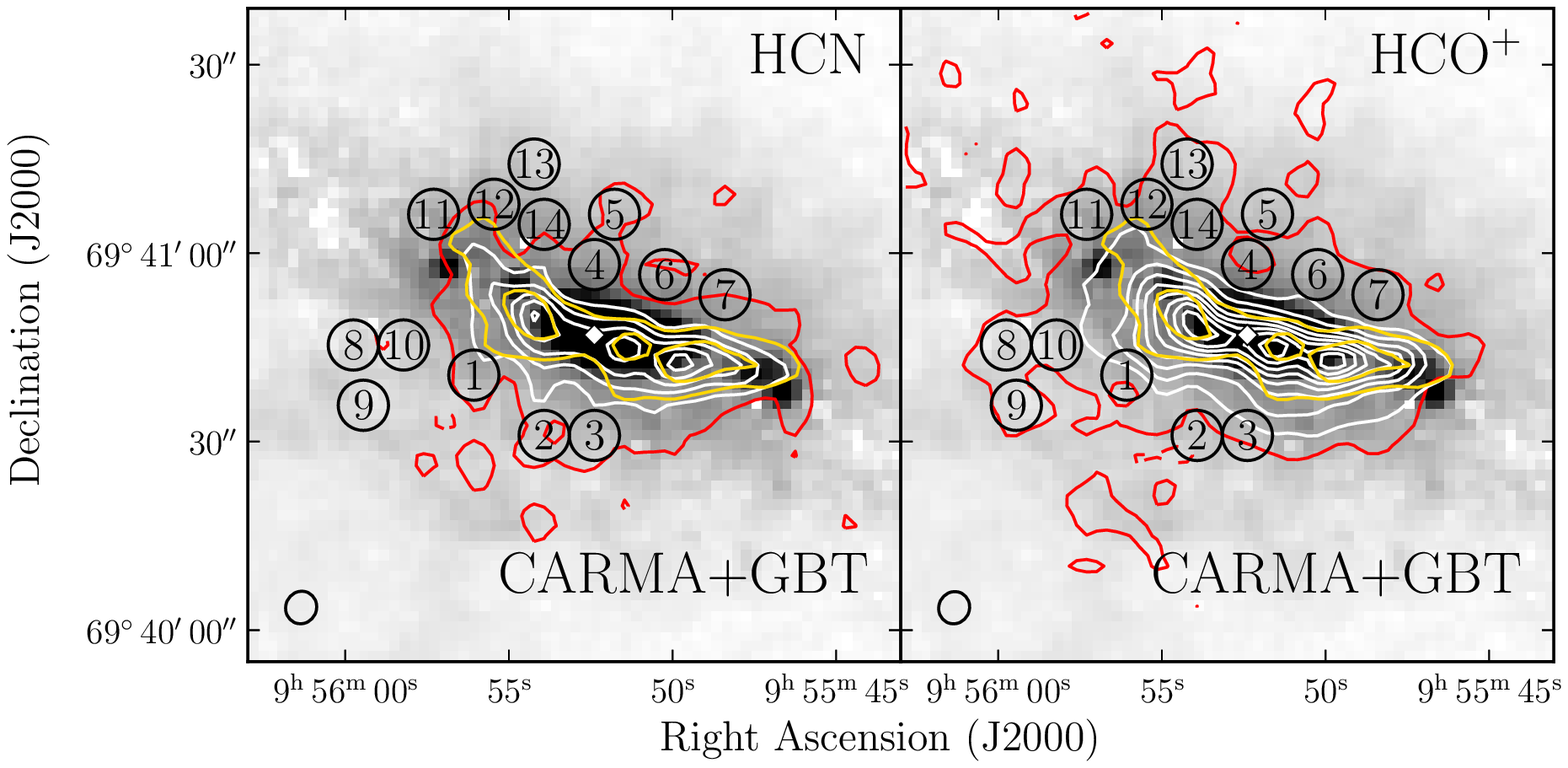}
\caption{\label{fig:mom0_ssc}
Short spacings-corrected integrated intensity (moment $0$) maps of HCN and \hco emission (solid contours)
overlaid on a map of \h2 $v=1-0$ S(1) emission \citep[inverted gray scale,][]{Veilleux2009}. Emission associated
with the central starburst is shown in \textit{white} and that associated with the starburst-driven wind is shown in
\textit{red}. The CO molecular disk is shown as yellow contours \citep{Walter2002} and the galaxy center defined by the
$2.2$~\mum\ peak as a white diamond \citep{Dietz1989}. The CARMA beam is shown in the lower left corner of each panel.
Contours start at $3\sigma$ and increment by $6\sigma$, with $\sigma$ the moment $0$ map rms.
{$\sigma_{\mathrm{HCN}}=2.5$~\Kkms}, {$\sigma_{\mathrm{\hco}}=2.5$~\Kkms}.
}
\end{figure*}

% \noindent
Using the velocity ranges defined above for the disk and extraplanar emission we construct moment $0$ maps using a
masked moment method as described in \citet{Dame2011}. In this method the data cubes are first convolved with a
larger beam to smooth out small scale variations. Then a threshold is applied to the smoothed data to create a mask
that is used to construct the moment $0$ maps.
For the convolution we use a Gaussian twice the \CARMA\ beam ($11\arcsec$) in the RA-DEC plane and two times the
velocity resolution ($21$~\kms) in the velocity axis. We use a threshold of $2.5$ times the convolved map rms to
recover emission. This method is good at recovering faint emission that is distributed similarly to the bright emission,
but is biased against isolated faint emission \citep{Regan2001,Helfer2003}. The moment $0$ maps are presented in
Figs.~\ref{fig:mom0_carma} and \ref{fig:mom0_ssc}, for the \CARMA\ data alone as well as the cubes corrected for
missing short spacings. In the moment $0$ maps we apply a $3\sigma$ cut to separate emission from noise. The 
differences between the short spacings-corrected data and the interferometric data alone are evident and we caution the
reader not to directly compare both. A comparison based on the \CARMA\ data alone shows that the \hco\ emission is the
most extended outside the plane, then follows HCN. Emission from HCN and \hco\ is slightly more extended than that of
the $3$~mm continuum (Fig.~\ref{fig:cont}). In Fig.~\ref{fig:mom0_ssc} we also include $14$ regions which are later
used to estimate the amount of molecular gas mass outside the central starburst. These regions have sizes of $7\farcs5$,
chosen to sample regions slightly larger than the synthesized beam. These regions are located where the extraplanar
emission extends at least $5\arcsec$ above or below the central starburst.

% \noindent
The short spacings-corrected moment $0$ maps show the presence of the extended component (Fig.~\ref{fig:mom0_ssc}).
Near the edge of the molecular disk the HCN and \hco\ emission show a chimney like structure. Similar to that observed
in [NeII] \citep{Achtermann1995} and $5$~GHz continuum \citep{Wills1999}. The extended \hco emission is coincident
with the dark lane in the Eastern edge of the $3$~mm continuum in Fig.~\ref{fig:cont} \citep[also in
Fig.~$5$ of][]{Ohyama2002}.

% \noindent
The HCN and \hco\ emission are not spatially coincident in various regions outside the central starburst. In regions
$1$, $2$, $4$ and $5$ emission from \hco\ is not observed. The noise level in these regions on both maps are
similar. This implies that the observed differences are due to different emission levels. If we take the $3\sigma$ limit
for the missing molecule, the non-detection of \hco\ emission implies HCN/\hco\ ratios $\geq1.5$. The 
non-detection of HCN emission in regions $8$, $9$ and $10$ implies HCN/\hco\ ratios $\leq0.75$.

% \noindent
The most extended component observed in HCN reaches a projected distance relative to the $2.2$~\mum\ peak of
${\approx22\arcsec}$, just above region $5$ in Fig.~\ref{fig:mom0_ssc}. To the south of region $2$ the projected
distance is ${\approx21\arcsec}$. For \hco\ the most extended emission comes from a projected distance of
$\approx31\arcsec$, to the north of region $14$, and ${\approx37\arcsec}$, to the south of region $9$. The prominent
extra-planar emission from \hco\ (regions $9$ and $14$ in Fig.~\ref{fig:mom0_ssc}) is spatially coincident with the dust
lane that passes near the center of M82 (see Fig.~\ref{fig:cont}).

\subsection{Other detected molecules}

CS, HNC and \hc3n emission comes from within the molecular disk. \hc3n is detected in the North-East lobe, where
transitions up to $J=28$ have been detected from the same molecule \citep{Aladro2011b}. The lack of CS and HNC emission
outside the molecular disk could be due to a lower signal-to-noise ratio, as their emission is fainter than that of HCN
and \hco.

\subsection{SFR from $3$~mm continuum}

The $3$~mm continuum in M82 is $\geq60\%$ due to free-free emission \citep{Carlstrom1991}. Using the observed
$3$~mm continuum we derive a star formation rate (SFR) from \citep[e.g.,][]{Murphy2011}
\begin{equation}
 \begin{split}
  \left(\frac{\mbox{SFR}^{\rm{ff}}}{\Msunpyr}\right)=4.6\times10^{-28}\times\\
\left(\frac{T_{e}}{10^{4}\mbox{ K}}\right)^{-0.45}\left(\frac{\nu}{\mbox{GHz}}\right)^{0.1}\left(\frac{L_{\nu}^{\rm{ff}}
}{\ergphzps}\right).
 \end{split}
\end{equation}
Here $T_{e}$ is the electron temperature and $L_{\nu}^{\rm{ff}}$ the luminosity due to free-free emission. We adopt
an electron temperature of $5000$~K, derived from radio recombination lines observations \citep{Puxley1989}. If we
consider that the contribution to free-free emission is $60-100\%$ \citep{Carlstrom1991}, then the SFR rate is
$4-7$~\Msunpyr. This value is $20\%$ lower than the IR luminosity derived SFR presented by \citet{Strickland2004}.
The lower value presented here could be attributed to the smaller region being probed with the $3$~mm continuum.

\subsection{Dense molecular gas mass outside the central starburst}

To determine the column density we use the regions shown in Fig.~\ref{fig:mom0_ssc}. The area-averaged
velocity-integrated brightness temperatures inside these regions are given in Table~\ref{table:tbint}.
Velocity-integrated line fluxes over the whole molecular disk and outflow are given in Table~\ref{table:flux}.
For the outflow a velocity range of $115-367$~km~s$^{-1}$ is used, and for the molecular disk
$50-390$~km~s$^{-1}$.

% \noindent
We cannot determine the optical depth nor the excitation temperature of the gas as we lack information about
higher J transitions, hyperfine structure transitions or isotopic line ratios for the observed molecules we cannot
determine the optical depth nor the excitation temperature of the gas. Therefore we assume that the gas is
optically thin and in local thermodynamic equilibrium (LTE). While this assumption is not likely to hold in the
central starburst region of M82, it might be appropriate for the gas in a wind \citep{Weiss2005}. 

The overall effect of the optically thin and LTE assumptions is that the computed column densities are lower
limits. Studies using multiple transitions of HCN and \hco\ suggests optical depths close to unity
\citep[$\tau\sim1$,][]{Nguyen1992,Naylor2010} in the central starburst. A greater than unity optical depth would
increase the column density. Deviations from LTE are expected if the density is lower than its critical value. The
conditions in the outflow are similar to those of the low excitation component in the central starburst
\citep{Weiss2005}. 
The density in the outflow inferred from CO studies is $\approx10^{3}$~cm$^{-3}$, which would imply that the HCN and
\hco\ lines are sub-thermally excited. In the case of sub-thermal excitation the LTE assumption provides a lower
limit to the true column density. 

% \noindent
Under the optically thin and LTE assumptions the column density, $N$, is given by ~\citep[e.g.,][]{Nguyen1989}
\begin{equation}
N=\frac{3k_{\mathrm{B}}}{8\pi^{3}\nu\mu^{2}S}\frac{Q(T_{\mathrm{ex}})}{e^{-E_{\mathrm{u}}/T_{\mathrm{ex}}}}\int
T_{\mathrm{b}}dv.
\end{equation}
\noindent
Here $k_{\mathrm{B}}$ is the Boltzmann constant, $\nu$ the transition frequency, $\mu$ the molecule permanent dipole
moment, $S$ the line strength, $Q$ the partition function, $T_{\mathrm{ex}}$ the excitation temperature,
$E_{\mathrm{u}}$ the upper state energy and $T_{\mathrm{b}}$ the line brightness temperature in temperature units. 
We adopt $\mu^{2}S$ to be $8.91135$~D$^{2}$ for HCN \citep{Remijan2007} and $15.21$~D$^{2}$ for \hco\
\citep{Yamaguchi1994}.

\noindent
The second major uncertainty in our column density calculation comes from the poorly known gas temperature. Multi
transition analysis of the physical conditions in the molecular disk of M82 suggests that the excitation temperature of
the HCN and \hco\ emitting gas is greater than $40$~K \citep{Naylor2010}. Outside the molecular disk, the gas
temperatures seem to be above $30$~K and below $120$~K \citep{Weiss2005}. We adopt a temperature range of $30-120$~K to
compute the column densities. The column density changes by $70\%$ between the extremes of this temperature range.
The column densities derived under these assumptions are presented in Table~\ref{table:cdens}. As the interferometer
beam is not able to resolve individual clouds, the values in Table~\ref{table:cdens} are {\bf area} averaged column
densities $\langle N\rangle$, the product between column density and area filling factor.

\begin{deluxetable}{lcc}
\tablecaption{Average $\int T_{b}dv$ outside the central starburst in M82.\label{table:tbint}}

\tablehead{
\colhead{Region} & \colhead{$\int T_{b,\hcn}dv$} & \colhead{$\int T_{b,\hco}dv$} \\
\colhead{} & \colhead{(\Kkms)} & \colhead{(\Kkms)}
}

\startdata
1  & $15\pm4$ & $24\pm11$\tablenotemark{$\dagger$} \\
2  & $11\pm2$ & $10.0\pm0.2$\tablenotemark{$\dagger$} \\
3  & $14\pm3$ & $19\pm8$\tablenotemark{$\dagger$} \\
4  & $14\pm3$ & $12\pm3$\tablenotemark{$\dagger$} \\
5  & $12\pm3$ & $9\pm0.3$\tablenotemark{$\dagger$} \\
6  & $8\pm1$\tablenotemark{$\dagger$} & $13\pm3$ \\
7  & $9\pm2$\tablenotemark{$\dagger$} & $16\pm2$ \\
8  & \nodata  & $13\pm1$ \\
9  & \nodata  & $10\pm1$ \\
10 & \nodata  & $15\pm2$ \\
11 & $8\pm2$\tablenotemark{$\dagger$} & $14\pm3$ \\
12 & $9\pm2$\tablenotemark{$\dagger$} & $17\pm4$ \\
13 & \nodata  & $14\pm1$ \\
14 & $10\pm1$\tablenotemark{$\dagger$} & $13\pm2$ 
\enddata

\tablenotetext{$\dagger$}{Regions filled less than $50\%$.}

\end{deluxetable}

\begin{deluxetable}{lcc}

\tablecaption{Integrated line fluxes in M82\label{table:flux}}

\tablehead{
\colhead{Region} & \colhead{$S_{\mathrm{HCN}}$} & \colhead{$S_{\hco}$} \\
\colhead{} & \colhead{(\Jykms)} & \colhead{(\Jykms)}
}

\startdata
Disk\tablenotemark{a} & $192\pm4$ & $365\pm7$ \\
Outflow\tablenotemark{b} & $55\pm3$ & $112\pm8$
\enddata

\tablenotetext{a}{Disk area is $676$ $(\arcsec)^{2}$ for HCN and $1011$ $(\arcsec)^{2}$ for \hco.}
\tablenotetext{b}{Outflow area for HCN is $940$ $(\arcsec)^{2}$ and for \hco\ $1661$ $(\arcsec)^{2}$.}

\end{deluxetable}

\begin{deluxetable}{lcc}

\tablecaption{Lower limits to the column density outside the central starburst in M82.\label{table:cdens}}

\tablehead{
\colhead{Region} & \colhead{$\langle N_{\hcn}\rangle$} & \colhead{$\langle N_{\hco}\rangle$} \\ [+.5ex]
\colhead{} & \colhead{($\times10^{13}$ cm$^{-2}$)} & \colhead{($\times10^{13}$ cm$^{-2}$)} 
}

\startdata
1  & $5.2-18.7$ & $4.9-17.3$\tablenotemark{$\dagger$} \\
2  & $3.9-13.8$ & $2.0-7.2$\tablenotemark{$\dagger$} \\
3  & $4.9-17.6$ & $3.9-13.9$\tablenotemark{$\dagger$} \\
4  & $5.0-17.8$ & $2.4-8.6$\tablenotemark{$\dagger$} \\
5  & $4.5-15.9$ & $1.8-6.5$\tablenotemark{$\dagger$} \\
6  & $2.9-10.5$\tablenotemark{$\dagger$} & $2.6-9.2$ \\
7  & $3.3-11.8$\tablenotemark{$\dagger$} & $3.3-11.5$ \\
8  & \nodata  & $2.8-9.9$ \\
9  & \nodata  & $2.2-7.7$ \\
10 & \nodata  & $3.1-11.0$ \\
11 & $2.8-9.9$\tablenotemark{$\dagger$} & $2.8-9.9$ \\
12 & $3.2-11.3$\tablenotemark{$\dagger$} & $3.5-12.4$ \\
13 & \nodata & $2.8-9.8$ \\
14 & $3.7-13.0$\tablenotemark{$\dagger$} & $2.7-9.5$
\enddata

% \tablenotetext{${a}$}{Listed values are lower limits.}
\tablenotetext{$\dagger$}{Regions filled less than $50\%$.}

\end{deluxetable}

% \noindent
Column densities are converted to molecular gas mass using \citep[see e.g.,][]{Kamentzky2012}
\begin{equation}
M_{\rm{region}}=\frac{m_{\h2}A_{\rm{reg}}N_{\rm{mol}}\Phi_{A}}{X_{\rm{mol}}},
\end{equation}
were $A_{\rm{reg}}$ is the area of the region being considered, $\Phi_{A_{\rm{reg}}}$ is the area filling factor of
the region, $m_{\h2}$ is the mass of an \h2\ molecule, $N_{\rm{mol}}$ is the column density of the molecule being used
as mass tracer and $X_{\rm{mol}}$ abundance of this molecule with respect to \h2
(${X_{\rm{mol}}=N_{\rm{mol}}/N_{\rm{H}_{2}}}$). For the HCN and \hco\ abundances we take the values derived by
\citet{Naylor2010} of ${X_{\hcn}\approx7.9\times10^{-9}}$ and ${X_{\hco}\approx3.9\times10^{-9}}$, which represent an
average of the abundances in the central starburst. The computed dense molecular gas masses are presented in
Table~\ref{table:mass}. The lower limits are of a few $\times10^{6}$~\Msun, similar to the mass of a giant molecular
cloud. As comparison, the mass of molecular gas inside the central starburst is $\geq9.8\times10^{7}$~\Msun\ for HCN and
$\geq14\times10^{7}$~\Msun\ for \hco, assuming optically thin emission in LTE. From this we can see that the
molecular gas mass in the outflow is a factor of $10$ lower than the mass in the disk. From CO observations, the mass
of molecular gas in the disk is $2.3\times10^{8}$~\Msun \citep{Walter2002}. In the disk, the gas masses traced by CO,
HCN and \hco\ seem to agree reasonably well with each other, given the large uncertainties and the differences in
critical density. In the outflow the molecular mass traced by CO($1-0$) is $3.3\times10^{8}$~\Msun
\citep{Walter2002}. The molecular gas mass traced by HCN and \hco\ is $\gtrsim2\%$ of the total molecular mass in the
outflow traced by CO.

\begin{deluxetable}{lcc}

\tablecaption{Lower limits to the dense molecular gas mass outside the central starburst in
M82.\label{table:mass}}

\tablehead{
\colhead{Region} & \colhead{$M_{\h2}$ from HCN} & \colhead{$M_{\h2}$ from \hco} \\
\colhead{} & \colhead{($\times10^{6}$ \Msun)} & \colhead{($\times10^{6}$ \Msun)} 
}

\startdata
1     & $1.5-5.5$ & $2.0-7.3$\tablenotemark{$\dagger$} \\
2     & $0.8-2.9$ & $0.2-0.6$\tablenotemark{$\dagger$} \\
3     & $1.5-5.2$ & $1.5-5.5$\tablenotemark{$\dagger$} \\
4     & $2.2-7.8$ & $0.7-2.7$\tablenotemark{$\dagger$} \\
5     & $1.0-3.6$ & $0.2-0.5$\tablenotemark{$\dagger$} \\
6     & $0.2-0.7$\tablenotemark{$\dagger$} & $1.5-5.4$ \\
7     & $0.7-2.5$\tablenotemark{$\dagger$} & $1.9-6.7$ \\
8     & \nodata  & $1.6-5.8$ \\
9     & \nodata  & $1.3-4.5$ \\
10    & \nodata  & $1.8-6.5$ \\
11    & $0.2-0.7$\tablenotemark{$\dagger$} & $1.6-5.8$ \\
12    & $0.4-1.6$\tablenotemark{$\dagger$} & $1.7-5.9$ \\
13    & \nodata  & $1.6-5.7$ \\
14    & $0.2-0.7$\tablenotemark{$\dagger$} & $1.6-5.6$ \\
total & $7.0-25.0$ & $14.6-51.9$
\enddata

% \tablenotetext{${a}$}{Listed values are lower limits.}
\tablenotetext{$\dagger$}{Regions filled less than $50\%$. These are not considered to compute the total mass.}
\end{deluxetable}

\subsection{Kinematics}

\begin{figure*}
\includegraphics[width=1\linewidth,angle=0]{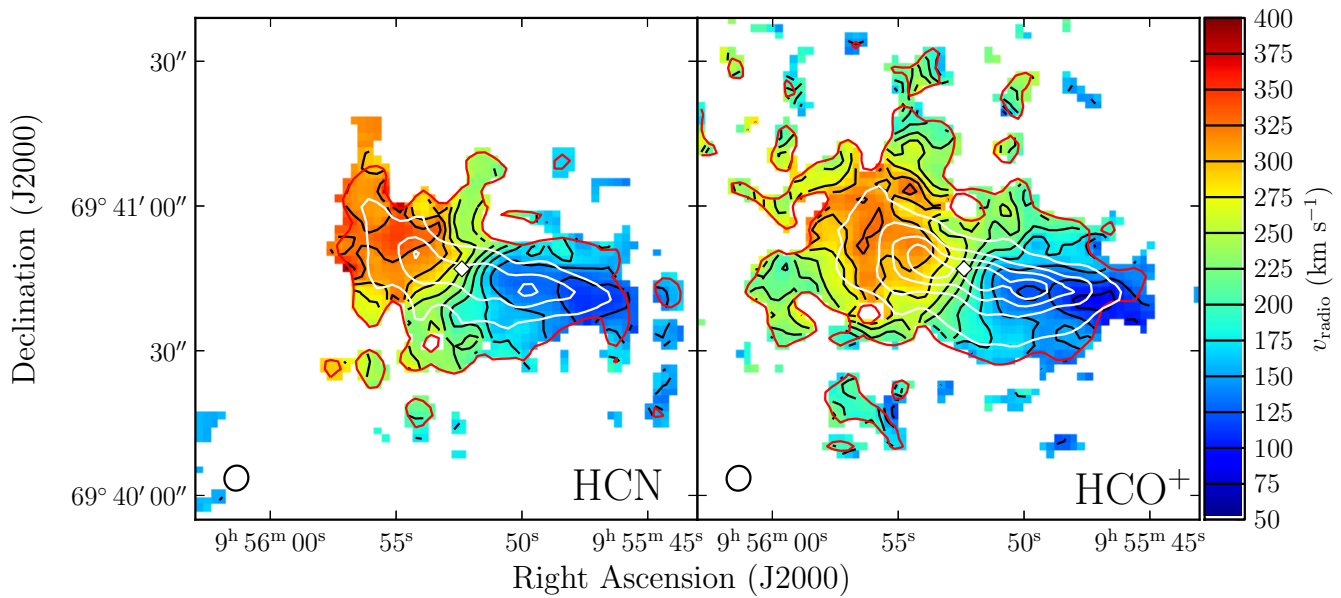}
\caption{\label{fig:mom1_ssc}
Line of sight velocity (moment $1$) maps of line emission in the central starburst of M82 (black contours and
color scale). The galaxy center defined by the $2.2$~\mum\ peak is shown as a white diamond \citep{Dietz1989}. The CARMA
beam is shown in the lower left corner of each panel. \textit{White} contours represent the moment $0$ map and increment
by $12\sigma$.}
\end{figure*}

Moment $1$ maps of the short spacings-corrected cubes are shown in Fig.~\ref{fig:mom1_ssc}. The moment $1$ was obtained
using the same masked moment technique as for the moment $0$ map. The velocity of the prominent \hco\ emission
extending south (regions $8$, $9$ and $10$) and north (regions $12$, $13$ and $14$) of the central starburst is close to
$232$~\kms\ and $263$~\kms\ respectively. After subtracting the systematic velocity of M82 ($220$~\kms) the velocities
are $12$ and $43$~\kms. The wind deprojected velocity can be obtained by correcting for the outflow opening angle
\citep[$55\arcdeg$,][]{Walter2002} and the galaxy inclination \citep[$80\arcdeg$,][]{deVaucouleurs1991}. This results in
deprojeted outflow velocities of ${\approx28}$ and ${\approx100}$~\kms. We adopt the average velocity between them,
$64$~\kms, as the \hco\ outflow velocity.

% \noindent
For the HCN emitting gas extending north (region $5$) and south (region $2$) of the central starburst, the observed
velocities are the same, ${\approx238}$~\kms, given our velocity resolution. This lack of velocity difference could be
due to the distribution of the HCN emitting gas in the outflow. The derived deprojected outflow velocity for HCN is
${\approx43}$~\kms.

% \noindent
The observed velocities for the HCN and \hco\ emitting gas do not show a blueshifted component, as observed at
other wavelengths \citep[see e.g.,][]{McKeith1995}. This is consistent with emission coming from near the galactic disk
in the part of the outflow closer to the observer to the north, and that farther from the observer to the south
\citep[regions II and IV in Fig.~$5$ of][]{McKeith1995}. Also, it is possible that the velocity of the dense molecular
gas is reflecting the circular velocity it had when it was part of the molecular disk \citep{Murray2011}.

\subsection{Dense gas mass outflow rate}

The time it takes to the dense gas to reach its observed position outside the central starburst can be estimated
from its deprojected velocity and distance from the molecular disk. The HCN emission from the outflow reaches a
projected distance of $21\arcsec$, which corresponds to a deprojected linear distance of $435$~pc. The \hco\ filament to
the North reaches a deprojected distance of $643$~pc, and to the South of $767$~pc. Given these distances and the
observed gas velocity, it would have taken $\sim14$~Myr for the HCN and \hco\ gas to reach its current location. This
timescale is comparable to the age of the most recent starburst in M82 A of $5$~Myr \citep{Forster2003}, and also
similar to the inferred lifetime of the SiO chimney and supershell of $\sim1$~Myr \citep{Garcia2001}.

% \noindent
Using the derived values of the dense molecular gas mass in the outflow, and the time required for the gas to reach the
farthest point from the $2.2$ \mum\ peak we can estimate the mass outflow rate for the dense gas, given by
${\dot{M}\approx M/t}$. We consider the most clear examples of outflowing dense molecular gas, regions $2$, $4$ and $5$
for HCN, and regions $8$ to $10$ and $12$ to $14$ for \hco. The total dense molecular gas mass in the mentioned
HCN regions is ${\geq4\times10^{6}}$~\Msun. For \hco\ this is ${\geq9.6\times10^{6}}$~\Msun.
The dense gas mass outflow rates are then ${\sim0.3}$~\Msunpyr\ for HCN and ${\sim0.7}$~\Msunpyr\ for \hco. These mass
outflow rates are lower than the SFR derived from the $3$~mm continuum and the mass loss rate inferred from CO
observations of $33$~\Msunpyr\ \citep{Walter2002}.

% \noindent
To the East of the central starburst there is a filament of \hco\ emission. This filament originates in region $11$
of Fig.~\ref{fig:mom0_ssc}. The velocity of this filament is lower than that of the molecular disk. This is similar to
what is observed for the CO emitting gas in this region \citep{Walter2002}.

\subsection{Energetics}

Using the derived values for the outflow mass and velocity we can estimate the kinetic energy of the dense molecular gas
outside the molecular disk. For HCN this is ${\approx4.7\times10^{52}}$~erg, and for \hco\
${\approx3.2\times10^{53}}$~erg. The source of this energy could be easily provided by the SNe explosions in M82.
The energy injected to the ISM can be estimated as ${\eta_{\rm{inj}}\dot{N}_{\rm{SNe}}E_{\rm{SN}}\Delta t}$, were
$\eta_{\rm{inj}}$ is the efficiency with which a SNe transfers energy to the ISM, $\dot{N}_{\rm{SNe}}$ is the observed
supernova rate, $E_{\rm{SN}}$ is the energy of a typical SNe and $\Delta t$ a time interval. We assume that each SNe
releases an energy of $10^{51}$~erg and that only $10\%$ of it is transfered to the surrounding gas
\citep[e.g.,][]{Thornton1998}. For the SNe rate we use a value of $0.04$~SNe~yr$^{-1}$ inferred from the observed SNe
remnants (SNR) in the starburst disk {\citep{Fenech2010}}. Over a timescale of $1$~Myr this would supply
${4\times10^{55}}$~erg. The derived outflow energy is $0.1\%$ of the energy provided by SNe under these assumptions.
This could be caused by a lower SNe rate in the region, or a lower SNe energy injection efficiency to the dense gas.

% \noindent
The derived kinetic energy of the dense molecular gas is lower than that of the less dense gas traced by CO
\citep{Walter2002} by a factor of $10^{2}-10^{3}$. The situation is similar with respect to the kinetic energy of
the ionized gas and the neutral atomic gas in the ouflow. The ionized gas traced by H$\alpha$ has a kinetic energy of
${2\times10^{55}}$~erg \citep{Shopbell1998} and the neutral atomic gas of ${(1-5)\times10^{54}}$~erg
{\citep{Contursi2013}}. The kinetic energy of the warm molecular gas is $10^{51}$~erg \citep{Veilleux2009}. This
makes the dense cool molecular gas one of the least energetic components of the outflow.

\section{Summary}
\label{sec:summary}

We presented HCN and \hco\ observations of the starburst galaxy M82. These observations are a combination of high
sensitivity single dish images obtained with the GBT, and high resolution CARMA observations. They cover a region of
$1.2$~kpc and trace structures of size ${\approx100}$~pc, including the low surface brightness emission associated
with the starburst-driven outflow.

% \noindent
The distribution and kinematics of the HCN and \hco\ emission suggests that the gas is being blown out of the central
molecular disk through at least one chimney, similar to what is observed at other wavelengths in M82. 
Assuming that the gas in the outflow is in LTE and optically thin we place lower limits to the mass of dense molecular
gas in the outflow. Using the HCN($1-0$) and HCO$^{+}$($1-0$) lines the mass of dense gas is
${\gtrsim7\times10^{6}}$~\Msun. This is $\geq2\%$ of the total outflow mass as traced by the CO $\mbox{J}=1-0$ line.

The energy required to expel this amount of dense gas from the central starburst is $(1-10)\times10^{52}$~erg. This
energy is about $0.1\%$ of the total energy provided by SNe in M82.

The rate at which dense gas is being blown out of the central starburst would imply that the starburst episode
would last $\geq5\%$ less than in the absence of the outflow of dense molecular gas.

\section*{Acknowledgments}

We thank the anonymous referee for providing feedback which helped improve the content and clarity of this work.
G.~G. and P.~S. acknowledge support from proyecto FONDECYT regular $1120195$, proyecto Anillo ACT$1102$
and Basal PFB-06 CATA.
A.~D.~B. wishes to acknowledge partial support from a CAREER grant NSF-AST$0955836$, NSF-AST$1139998$ and from
NSF-AST$1412419$. This research made use of Astropy \citep{Astropy2013}. \\

\bibliographystyle{apj}
% \bibliography{m82}

\begin{thebibliography}{}

\bibitem[{{Achtermann} \& {Lacy}(1995)}]{Achtermann1995}
{Achtermann}, J.~M., \& {Lacy}, J.~H. 1995, \apj, 439, 163

\bibitem[{{Aladro} {et~al.}(2011b){Aladro}, {Mart{\'{\i}}n},
  {Mart{\'{\i}}n-Pintado}, {Mauersberger}, {Henkel}, {Oca{\~n}a Flaquer}, \&
  {Amo-Baladr{\'o}n}}]{Aladro2011b}
{Aladro}, R., {Mart{\'{\i}}n}, S., {Mart{\'{\i}}n-Pintado}, J., {et~al.} 2011b,
  \aap, 535, A84

\bibitem[{{Aladro} {et~al.}(2011a){Aladro}, {Mart{\'{\i}}n-Pintado},
  {Mart{\'{\i}}n}, {Mauersberger}, \& {Bayet}}]{Aladro2011a}
{Aladro}, R., {Mart{\'{\i}}n-Pintado}, J., {Mart{\'{\i}}n}, S., {Mauersberger},
  R., \& {Bayet}, E. 2011a, \aap, 525, A89

\bibitem[{{Astropy Collaboration} {et~al.}(2013)}]{Astropy2013}
{Astropy Collaboration}, {et~al.} 2013, \aap, 558, A33

\bibitem[{{Bolatto} {et~al.}(2013){Bolatto}, {Warren}, {Leroy}, {Walter},
  {Veilleux}, {Ostriker}, {Ott}, {Zwaan}, {Fisher}, {Weiss}, {Rosolowsky}, \&
  {Hodge}}]{Bolatto2013}
{Bolatto}, A.~D., {Warren}, S.~R., {Leroy}, A.~K., {et~al.} 2013, \nat, 499,
  450

\bibitem[{{Bregman} {et~al.}(1995){Bregman}, {Schulman}, \&
  {Tomisaka}}]{Bregman1995}
{Bregman}, J.~N., {Schulman}, E., \& {Tomisaka}, K. 1995, \apj, 439, 155

\bibitem[{{Brouillet} \& {Schilke}(1993)}]{Brouillet1993}
{Brouillet}, N., \& {Schilke}, P. 1993, \aap, 277, 381

\bibitem[{{Carlstrom}(1989)}]{Carlstrom1989}
{Carlstrom}, J.~E. 1989, PhD thesis, California Univ., Berkeley.

\bibitem[{{Carlstrom} \& {Kronberg}(1991)}]{Carlstrom1991}
{Carlstrom}, J.~E., \& {Kronberg}, P.~P. 1991, \apj, 366, 422

\bibitem[{{Contursi} {et~al.}(2013){Contursi}, {Poglitsch}, {Gr{\'a}cia
  Carpio}, {Veilleux}, {Sturm}, {Fischer}, {Verma}, {Hailey-Dunsheath}, {Lutz},
  {Davies}, {Gonz{\'a}lez-Alfonso}, {Sternberg}, {Genzel}, \&
  {Tacconi}}]{Contursi2013}
{Contursi}, A., {Poglitsch}, A., {Gr{\'a}cia Carpio}, J., {et~al.} 2013, \aap,
  549, A118

\bibitem[{{Dalcanton} {et~al.}(2009){Dalcanton}, {Williams}, {Seth}, {Dolphin},
  {Holtzman}, {Rosema}, {Skillman}, {Cole}, {Girardi}, {Gogarten},
  {Karachentsev}, {Olsen}, {Weisz}, {Christensen}, {Freeman}, {Gilbert},
  {Gallart}, {Harris}, {Hodge}, {de Jong}, {Karachentseva}, {Mateo}, {Stetson},
  {Tavarez}, {Zaritsky}, {Governato}, \& {Quinn}}]{Dalcanton2009}
{Dalcanton}, J.~J., {Williams}, B.~F., {Seth}, A.~C., {et~al.} 2009, \apjs,
  183, 67

\bibitem[{{Dame}(2011)}]{Dame2011}
{Dame}, T.~M. 2011, ArXiv e-prints

\bibitem[{{de Vaucouleurs} {et~al.}(1991){de Vaucouleurs}, {de Vaucouleurs},
  {Corwin}, {Buta}, {Paturel}, \& {Fouqu{\'e}}}]{deVaucouleurs1991}
{de Vaucouleurs}, G., {de Vaucouleurs}, A., {Corwin}, Jr., H.~G., {et~al.}
  1991, {Third Reference Catalogue of Bright Galaxies. Volume I: Explanations
  and references. Volume II: Data for galaxies between 0$^{h}$ and 12$^{h}$.
  Volume III: Data for galaxies between 12$^{h}$ and 24$^{h}$.}

\bibitem[{{Dietz} {et~al.}(1989){Dietz}, {Gehrz}, {Jones}, {Grasdalen},
  {Smith}, {Gullixson}, \& {Hackwell}}]{Dietz1989}
{Dietz}, R.~D., {Gehrz}, R.~D., {Jones}, T.~J., {et~al.} 1989, \aj, 98, 1260

\bibitem[{{Engelbracht} {et~al.}(2006){Engelbracht}, {Kundurthy}, {Gordon},
  {Rieke}, {Kennicutt}, {Smith}, {Regan}, {Makovoz}, {Sosey}, {Draine},
  {Helou}, {Armus}, {Calzetti}, {Meyer}, {Bendo}, {Walter}, {Hollenbach},
  {Cannon}, {Murphy}, {Dale}, {Buckalew}, \& {Sheth}}]{Engelbracht2006}
{Engelbracht}, C.~W., {Kundurthy}, P., {Gordon}, K.~D., {et~al.} 2006, \apjl,
  642, L127

\bibitem[{{Fenech} {et~al.}(2010){Fenech}, {Beswick}, {Muxlow}, {Pedlar}, \&
  {Argo}}]{Fenech2010}
{Fenech}, D., {Beswick}, R., {Muxlow}, T.~W.~B., {Pedlar}, A., \& {Argo}, M.~K.
  2010, \mnras, 408, 607

\bibitem[{{F{\"o}rster Schreiber} {et~al.}(2003){F{\"o}rster Schreiber},
  {Genzel}, {Lutz}, \& {Sternberg}}]{Forster2003}
{F{\"o}rster Schreiber}, N.~M., {Genzel}, R., {Lutz}, D., \& {Sternberg}, A.
  2003, \apj, 599, 193

\bibitem[{{Garc{\'{\i}}a-Burillo} {et~al.}(2001){Garc{\'{\i}}a-Burillo},
  {Mart{\'{\i}}n-Pintado}, {Fuente}, \& {Neri}}]{Garcia2001}
{Garc{\'{\i}}a-Burillo}, S., {Mart{\'{\i}}n-Pintado}, J., {Fuente}, A., \&
  {Neri}, R. 2001, \apjl, 563, L27

\bibitem[{{Helfer} {et~al.}(2003){Helfer}, {Thornley}, {Regan}, {Wong},
  {Sheth}, {Vogel}, {Blitz}, \& {Bock}}]{Helfer2003}
{Helfer}, T.~T., {Thornley}, M.~D., {Regan}, M.~W., {et~al.} 2003, \apjs, 145,
  259

\bibitem[{{Jura} {et~al.}(1978){Jura}, {Hobbs}, \& {Maran}}]{Jura1978}
{Jura}, M., {Hobbs}, R.~W., \& {Maran}, S.~P. 1978, \aj, 83, 153

\bibitem[{{Kamenetzky} {et~al.}(2012){Kamenetzky}, {Glenn}, {Rangwala},
  {Maloney}, {Bradford}, {Wilson}, {Bendo}, {Baes}, {Boselli}, {Cooray},
  {Isaak}, {Lebouteiller}, {Madden}, {Panuzzo}, {Schirm}, {Spinoglio}, \&
  {Wu}}]{Kamentzky2012}
{Kamenetzky}, J., {Glenn}, J., {Rangwala}, N., {et~al.} 2012, \apj, 753, 70

\bibitem[{{Kepley} {et~al.}(2014){Kepley}, {Leroy}, {Frayer}, {Usero},
  {Marvil}, \& {Walter}}]{Kepley2014}
{Kepley}, A.~A., {Leroy}, A.~K., {Frayer}, D., {et~al.} 2014, \apjl, 780, L13

\bibitem[{{Lim} {et~al.}(2013){Lim}, {Hwang}, \& {Lee}}]{Lim2013}
{Lim}, S., {Hwang}, N., \& {Lee}, M.~G. 2013, \apj, 766, 20

\bibitem[{{Loiseau} {et~al.}(1990){Loiseau}, {Nakai}, {Sofue}, {Wielebinski},
  {Reuter}, \& {Klein}}]{Loiseau1990}
{Loiseau}, N., {Nakai}, N., {Sofue}, Y., {et~al.} 1990, \aap, 228, 331

\bibitem[{{Lynds} \& {Sandage}(1963)}]{Lynds1963}
{Lynds}, C.~R., \& {Sandage}, A.~R. 1963, \apj, 137, 1005

\bibitem[{{McKeith} {et~al.}(1995){McKeith}, {Greve}, {Downes}, \&
  {Prada}}]{McKeith1995}
{McKeith}, C.~D., {Greve}, A., {Downes}, D., \& {Prada}, F. 1995, \aap, 293,
  703

\bibitem[{{Melioli} {et~al.}(2013){Melioli}, {de Gouveia Dal Pino}, \&
  {Geraissate}}]{Melioli2013}
{Melioli}, C., {de Gouveia Dal Pino}, E.~M., \& {Geraissate}, F.~G. 2013,
  \mnras, 430, 3235

\bibitem[{{Murphy} {et~al.}(2011){Murphy}, {Condon}, {Schinnerer}, {Kennicutt},
  {Calzetti}, {Armus}, {Helou}, {Turner}, {Aniano}, {Beir{\~a}o}, {Bolatto},
  {Brandl}, {Croxall}, {Dale}, {Donovan Meyer}, {Draine}, {Engelbracht},
  {Hunt}, {Hao}, {Koda}, {Roussel}, {Skibba}, \& {Smith}}]{Murphy2011}
{Murphy}, E.~J., {Condon}, J.~J., {Schinnerer}, E., {et~al.} 2011, \apj, 737,
  67

\bibitem[{{Murray} {et~al.}(2011){Murray}, {M{\'e}nard}, \&
  {Thompson}}]{Murray2011}
{Murray}, N., {M{\'e}nard}, B., \& {Thompson}, T.~A. 2011, \apj, 735, 66

\bibitem[{{Mutchler} {et~al.}(2007){Mutchler}, {Bond}, {Christian}, {Frattare},
  {Hamilton}, {Januszewski}, {Levay}, {Mountain}, {Noll}, {Royle}, {Gallagher},
  \& {Puxley}}]{Mutchler2007}
{Mutchler}, M., {Bond}, H.~E., {Christian}, C.~A., {et~al.} 2007, \pasp, 119, 1

\bibitem[{{Nakai} {et~al.}(1987){Nakai}, {Hayashi}, {Handa}, {Sofue},
  {Hasegawa}, \& {Sasaki}}]{Nakai1987}
{Nakai}, N., {Hayashi}, M., {Handa}, T., {et~al.} 1987, \pasj, 39, 685

\bibitem[{{Naylor} {et~al.}(2010){Naylor}, {Bradford}, {Aguirre}, {Bock},
  {Earle}, {Glenn}, {Inami}, {Kamenetzky}, {Maloney}, {Matsuhara}, {Nguyen}, \&
  {Zmuidzinas}}]{Naylor2010}
{Naylor}, B.~J., {Bradford}, C.~M., {Aguirre}, J.~E., {et~al.} 2010, \apj, 722,
  668

\bibitem[{{Neininger} {et~al.}(1998){Neininger}, {Guelin}, {Klein},
  {Garcia-Burillo}, \& {Wielebinski}}]{Neininger1998}
{Neininger}, N., {Guelin}, M., {Klein}, U., {Garcia-Burillo}, S., \&
  {Wielebinski}, R. 1998, \aap, 339, 737

\bibitem[{{Nguyen} {et~al.}(1992){Nguyen}, {Jackson}, {Henkel}, {Truong}, \&
  {Mauersberger}}]{Nguyen1992}
{Nguyen}, Q.-R., {Jackson}, J.~M., {Henkel}, C., {Truong}, B., \&
  {Mauersberger}, R. 1992, \apj, 399, 521

\bibitem[{{Nguyen-Q-Rieu} {et~al.}(1989){Nguyen-Q-Rieu}, {Nakai}, \&
  {Jackson}}]{Nguyen1989}
{Nguyen-Q-Rieu}, {Nakai}, N., \& {Jackson}, J.~M. 1989, \aap, 220, 57

\bibitem[{{O'Connell} \& {Mangano}(1978)}]{OConnell1978}
{O'Connell}, R.~W., \& {Mangano}, J.~J. 1978, \apj, 221, 62

\bibitem[{{Ohyama} {et~al.}(2002){Ohyama}, {Taniguchi}, {Iye}, {Yoshida},
  {Sekiguchi}, {Takata}, {Saito}, {Kawabata}, {Kashikawa}, {Aoki}, {Sasaki},
  {Kosugi}, {Okita}, {Shimizu}, {Inata}, {Ebizuka}, {Ozawa}, {Yadoumaru},
  {Taguchi}, \& {Asai}}]{Ohyama2002}
{Ohyama}, Y., {Taniguchi}, Y., {Iye}, M., {et~al.} 2002, \pasj, 54, 891

\bibitem[{{Oppenheimer} \& {Dav{\'e}}(2006)}]{Oppenheimer2006}
{Oppenheimer}, B.~D., \& {Dav{\'e}}, R. 2006, \mnras, 373, 1265

\bibitem[{{Oppenheimer} {et~al.}(2010){Oppenheimer}, {Dav{\'e}}, {Kere{\v s}},
  {Fardal}, {Katz}, {Kollmeier}, \& {Weinberg}}]{Oppenheimer2010}
{Oppenheimer}, B.~D., {Dav{\'e}}, R., {Kere{\v s}}, D., {et~al.} 2010, \mnras,
  406, 2325

\bibitem[{{Puxley} {et~al.}(1989){Puxley}, {Brand}, {Moore}, {Mountain},
  {Nakai}, \& {Yamashita}}]{Puxley1989}
{Puxley}, P.~J., {Brand}, P.~W.~J.~L., {Moore}, T.~J.~T., {et~al.} 1989, \apj,
  345, 163

\bibitem[{{Regan} {et~al.}(2001){Regan}, {Thornley}, {Helfer}, {Sheth}, {Wong},
  {Vogel}, {Blitz}, \& {Bock}}]{Regan2001}
{Regan}, M.~W., {Thornley}, M.~D., {Helfer}, T.~T., {et~al.} 2001, \apj, 561,
  218

\bibitem[{{Remijan} {et~al.}(2007){Remijan}, {Markwick-Kemper}, \& {ALMA
  Working Group on Spectral Line Frequencies}}]{Remijan2007}
{Remijan}, A.~J., {Markwick-Kemper}, A., \& {ALMA Working Group on Spectral
  Line Frequencies}. 2007, in Bulletin of the American Astronomical Society,
  Vol.~39, American Astronomical Society Meeting Abstracts, \#132.11

\bibitem[{{Roussel} {et~al.}(2010){Roussel}, {Wilson}, {Vigroux}, {Isaak},
  {Sauvage}, {Madden}, {Auld}, {Baes}, {Barlow}, {Bendo}, {Bock}, {Boselli},
  {Bradford}, {Buat}, {Castro-Rodriguez}, {Chanial}, {Charlot}, {Ciesla},
  {Clements}, {Cooray}, {Cormier}, {Cortese}, {Davies}, {Dwek}, {Eales},
  {Elbaz}, {Galametz}, {Galliano}, {Gear}, {Glenn}, {Gomez}, {Griffin}, {Hony},
  {Levenson}, {Lu}, {O'Halloran}, {Okumura}, {Oliver}, {Page}, {Panuzzo},
  {Papageorgiou}, {Parkin}, {Perez-Fournon}, {Pohlen}, {Rangwala}, {Rigby},
  {Rykala}, {Sacchi}, {Schulz}, {Schirm}, {Smith}, {Spinoglio}, {Stevens},
  {Srinivasan}, {Symeonidis}, {Trichas}, {Vaccari}, {Wozniak}, {Wright}, \&
  {Zeilinger}}]{Roussel2010}
{Roussel}, H., {Wilson}, C.~D., {Vigroux}, L., {et~al.} 2010, \aap, 518, L66

\bibitem[{{Sanders} {et~al.}(2003){Sanders}, {Mazzarella}, {Kim}, {Surace}, \&
  {Soifer}}]{Sanders2003}
{Sanders}, D.~B., {Mazzarella}, J.~M., {Kim}, D.-C., {Surace}, J.~A., \&
  {Soifer}, B.~T. 2003, \aj, 126, 1607

\bibitem[{{Sault} {et~al.}(1995){Sault}, {Teuben}, \& {Wright}}]{Sault1995}
{Sault}, R.~J., {Teuben}, P.~J., \& {Wright}, M.~C.~H. 1995, in Astronomical
  Society of the Pacific Conference Series, Vol.~77, Astronomical Data Analysis
  Software and Systems IV, ed. R.~A. {Shaw}, H.~E. {Payne}, \& J.~J.~E.
  {Hayes}, 433

\bibitem[{{Sch{\"o}ier} {et~al.}(2005){Sch{\"o}ier}, {van der Tak}, {van
  Dishoeck}, \& {Black}}]{Schoier2005}
{Sch{\"o}ier}, F.~L., {van der Tak}, F.~F.~S., {van Dishoeck}, E.~F., \&
  {Black}, J.~H. 2005, \aap, 432, 369

\bibitem[{{Seaquist} {et~al.}(1998){Seaquist}, {Frayer}, \&
  {Bell}}]{Seaquist1998}
{Seaquist}, E.~R., {Frayer}, D.~T., \& {Bell}, M.~B. 1998, \apj, 507, 745

\bibitem[{{Shopbell} \& {Bland-Hawthorn}(1998)}]{Shopbell1998}
{Shopbell}, P.~L., \& {Bland-Hawthorn}, J. 1998, \apj, 493, 129

\bibitem[{{Springel} \& {Hernquist}(2003)}]{Springel2003}
{Springel}, V., \& {Hernquist}, L. 2003, \mnras, 339, 312

\bibitem[{{Stanimirovic}(2002)}]{Stanimirovic2002}
{Stanimirovic}, S. 2002, in Astronomical Society of the Pacific Conference
  Series, Vol. 278, Single-Dish Radio Astronomy: Techniques and Applications,
  ed. S.~{Stanimirovic}, D.~{Altschuler}, P.~{Goldsmith}, \& C.~{Salter},
  375--396

\bibitem[{{Stevens} {et~al.}(2003){Stevens}, {Read}, \&
  {Bravo-Guerrero}}]{Stevens2003}
{Stevens}, I.~R., {Read}, A.~M., \& {Bravo-Guerrero}, J. 2003, \mnras, 343, L47

\bibitem[{{Strickland} {et~al.}(2004){Strickland}, {Heckman}, {Colbert},
  {Hoopes}, \& {Weaver}}]{Strickland2004}
{Strickland}, D.~K., {Heckman}, T.~M., {Colbert}, E.~J.~M., {Hoopes}, C.~G., \&
  {Weaver}, K.~A. 2004, \apj, 606, 829

\bibitem[{{Strickland} {et~al.}(1997){Strickland}, {Ponman}, \&
  {Stevens}}]{Strickland1997}
{Strickland}, D.~K., {Ponman}, T.~J., \& {Stevens}, I.~R. 1997, \aap, 320, 378

\bibitem[{{Thornton} {et~al.}(1998){Thornton}, {Gaudlitz}, {Janka}, \&
  {Steinmetz}}]{Thornton1998}
{Thornton}, K., {Gaudlitz}, M., {Janka}, H.-T., \& {Steinmetz}, M. 1998, \apj,
  500, 95

\bibitem[{{Tsai} {et~al.}(2012){Tsai}, {Matsushita}, {Kong}, {Matsumoto}, \&
  {Kohno}}]{Tsai2012}
{Tsai}, A.-L., {Matsushita}, S., {Kong}, A.~K.~H., {Matsumoto}, H., \& {Kohno},
  K. 2012, \apj, 752, 38

\bibitem[{{Veilleux} {et~al.}(2005){Veilleux}, {Cecil}, \&
  {Bland-Hawthorn}}]{Veilleux2005}
{Veilleux}, S., {Cecil}, G., \& {Bland-Hawthorn}, J. 2005, \araa, 43, 769

\bibitem[{{Veilleux} {et~al.}(2009){Veilleux}, {Rupke}, \&
  {Swaters}}]{Veilleux2009}
{Veilleux}, S., {Rupke}, D.~S.~N., \& {Swaters}, R. 2009, \apjl, 700, L149

\bibitem[{{Walter} {et~al.}(2002){Walter}, {Weiss}, \& {Scoville}}]{Walter2002}
{Walter}, F., {Weiss}, A., \& {Scoville}, N. 2002, \apjl, 580, L21

\bibitem[{{Wei{\ss}} {et~al.}(2005){Wei{\ss}}, {Walter}, \&
  {Scoville}}]{Weiss2005}
{Wei{\ss}}, A., {Walter}, F., \& {Scoville}, N.~Z. 2005, \aap, 438, 533

\bibitem[{{Wills} {et~al.}(1999){Wills}, {Redman}, {Muxlow}, \&
  {Pedlar}}]{Wills1999}
{Wills}, K.~A., {Redman}, M.~P., {Muxlow}, T.~W.~B., \& {Pedlar}, A. 1999,
  \mnras, 309, 395

\bibitem[{{Yamagishi} {et~al.}(2012){Yamagishi}, {Kaneda}, {Ishihara}, {Kondo},
  {Onaka}, {Suzuki}, \& {Minh}}]{Yamagishi2012}
{Yamagishi}, M., {Kaneda}, H., {Ishihara}, D., {et~al.} 2012, \aap, 541, A10

\bibitem[{{Yamaguchi} {et~al.}(1994){Yamaguchi}, {Richards}, \&
  {Schaefer}}]{Yamaguchi1994}
{Yamaguchi}, Y., {Richards}, Jr., C.~A., \& {Schaefer}, III, H.~F. 1994, \jcp,
  101, 8945

\end{thebibliography}

\label{lastpage}

\end{document}